\documentclass[
preprint,%
secnumarabic,%
 amssymb, amsmath,%
 aip,cha,%
]{revtex4-1}

\usepackage{docs}%
\usepackage{bm}%
\usepackage[colorlinks=true,linkcolor=blue]{hyperref}%
\expandafter\ifx\csname package@font\endcsname\relax\else
 \expandafter\expandafter
 \expandafter\usepackage
 \expandafter\expandafter
 \expandafter{\csname package@font\endcsname}%
\fi
\usepackage{graphicx}
\hypersetup{citecolor=blue}

\begin{document}
\title{High transparency Bi$_2$Se$_3$ topological insulator nanoribbon Josephson junctions with low resistive noise properties}

\author{Gunta Kunakova}
\affiliation{Quantum Device Physics Laboratory, Department of Microtechnology and Nanoscience, Chalmers University of Technology, SE-41296 G\"oteborg, Sweden}
\affiliation{Institute of Chemical Physics, University of Latvia, Raina Blvd. 19, LV-1586, Riga, Latvia}

\author{Thilo Bauch}
\affiliation{Quantum Device Physics Laboratory, Department of Microtechnology and Nanoscience, Chalmers University of Technology, SE-41296 G\"oteborg, Sweden}

\author{Edoardo Trabaldo}
\affiliation{Quantum Device Physics Laboratory, Department of Microtechnology and Nanoscience, Chalmers University of Technology, SE-41296 G\"oteborg, Sweden}

\author{Jana Andzane}
\affiliation{Institute of Chemical Physics, University of Latvia, Raina Blvd. 19, LV-1586, Riga, Latvia}

\author{Donats Erts}
\affiliation{Institute of Chemical Physics, University of Latvia, Raina Blvd. 19, LV-1586, Riga, Latvia}

\author{Floriana Lombardi}
\email{floriana.lombardi@chalmers.se}
\affiliation{Quantum Device Physics Laboratory, Department of Microtechnology and Nanoscience, Chalmers University of Technology, SE-41296 G\"oteborg, Sweden}


\begin{abstract}
\begin{description}
\item[Abstract] Bi$_2$Se$_3$ nanoribbons, grown by catalyst-free Physical Vapour Deposition, have been used to fabricate high quality Josephson junctions with Al superconducting electrodes. The conductance  spectra (dI/dV) of the junctions  show clear dip-peak structures characteristic of multiple Andreev reflections. The temperature dependence of the  dip-peak features reveals a highly transparent Al/Bi$_2$Se$_3$ topological insulator nanoribbon interface and Josephson junction barrier. This is supported by the high values of the Bi$_2$Se$_3$ induced gap and of I$_c$R$_n$ (I$_c$ critical current, R$_n$ normal resistance of the junction) product both of the order of 160 $\mu$eV, a value  close to the Al gap. The devices present an extremely low relative resistance noise below 1$\times$10$^{-12}$ $\mu$m$^2$/Hz comparable to the best Al tunnel junctions, which indicates a high stability in the transmission coefficients of transport channels. The ideal Al/Bi$_2$Se$_3$ interface properties, perfect transparency for Cooper pair  transport in conjunction with low resistive noise make these junctions a suitable platform for further studies of the induced topological superconductivity and Majorana bound states physics.
\end{description}
\end{abstract}
\maketitle \section*{}
\indent The interest for hybrid Topological Insulator (TI) Josephson junctions has boosted after the prediction by Fu and Kane \cite{Fu2008} of an unconventional  chiral p$_x$ + $i$p$_y$ (p-wave) symmetry of the  proximity induced order parameter into the topological surface states. The chiral  induced p-wave is a prerequisite for the nucleation of localized Majorana states in a tri-junction geometry \cite{Fu2008} which are instrumental for topological quantum computation. In a multimode hybrid TI Josephson junction, with two terminal geometry, Majorana physics manifests as peculiar properties of a part of the Andreev bound states carrying the Josephson current: they give rise to an unconventional 4$\pi$ periodic current phase relation (CPR) coexisting with a 2$\pi$ periodic CPR of the conventional Andreev bound states \cite{Wiedenmann2016}. The relative weight between the 4$\pi$ and 2$\pi$ periodic Andreev bound states increases with the transparency of the junction  and in general by reducing the number of channels\cite{Snelder2013,Li2018a}. A direct way to achieve a low number of transport channels is to use TI with reduced dimensionality like very thin and narrow nanoribbons. Indeed quite recently, various theoretical proposals have shown the advantage to use Josephson junctions with TI nanoribbon, with suppressed bulk conduction, to realize Majorana fermions \cite{Manousakis2017}.\\
\indent In this letter we demonstrate the realization of Bi$_2$Se$_3$ nanoribbon Josephson junctions with Al electrodes with a) highly transparent Bi$_2$Se$_3$/Al interface, b) highly transparent barrier for Cooper pair transport and c) 
low resistance noise. The combination of all these properties are instrumental to study novel effects related to topological superconductivity.\\
\indent As already observed in Josephson devices made with 2-dimensional electron gases (2DEG) \cite{Chrestin1997, Kjaergaard2017}, the detection of Multiple Andreev Reflections (MAR) is a powerful tool to get information about the value of the induced gap $\Delta^\prime$ and insights into the Cooper pair transport. In situ growth of the superconducting electrodes on the 3D TI would be the ideal strategy to realize highly transparent Superconductor / 3D TI interfaces as demonstrated in the case of InAs nanowires and 2DEG \cite{Chang2015, Kjaergaard2017}. However at present, most of the results presented in literature, on the induced superconductivity in 3D TI's, refer to devices fabricated from flakes, exfoliated from single crystals or films, and then transferred to a substrate \cite{Cho2013, Galletti2014, Calvez2018}. In this case the interfaces with the superconductor are realized ex-situ after the removal of the oxide layer which unavoidably form on top of the 3D TI \cite{Kong2011a}.\\
\indent Here we report an excellent coupling between Bi$_2$Se$_3$ nanoribbons and Al, by using an ex-situ procedure for the definition of the interface. Differential conductance spectra display sub-gap structures resulting from MAR.\\
\begin{figure*}
    \centering
    \includegraphics[width=1.0\linewidth]{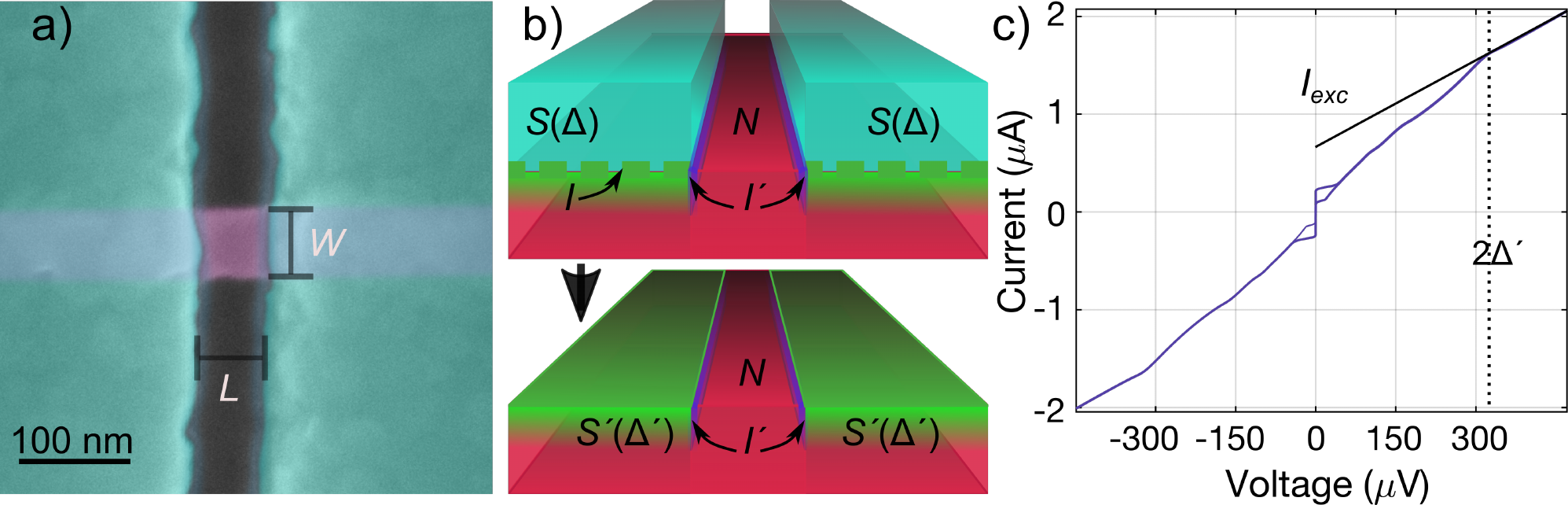}
    \caption{(a) Coloured SEM image of fabricated Bi$_2$Se$_3$ nanoribbon Josephson junction (in blue Al, violet Bi$_2$Se$_3$). (b) Schematic representation of a planar 3D TI Josephson junction. The superconducting electrodes (S) that induce superconductivity in the 3D TI underneath, indicated as S$^\prime$, are shown in blue. The green dashed line indicates the barrier I between S and S$^\prime$ while the dark blue region shows the barrier I$^\prime$ between S$^\prime$ and N representing the part of the 3D TI not covered by the electrodes. For large enough transparency between the 3D TI and Al one can model the system as a planar 2D junction (lower panel). (c) IVC of the junction B45-C1 measured at 20 mK. The black solid line is a linear extrapolation to zero voltage used to extract the excess current I$_{exc}$ while the black dotted line indicate value of twice the induced gap $\Delta^\prime$.}
    \label{DeviceSchem}
\end{figure*}
\begin{figure}
    \centering
    \includegraphics[width=0.5\linewidth]{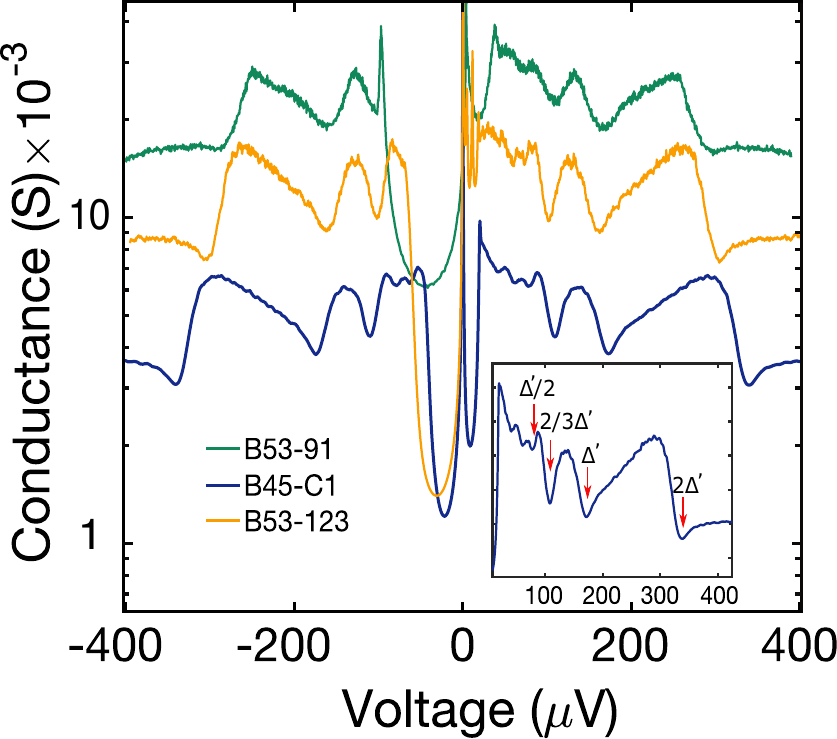}
    \caption{Conductance spectra (dI/dV) for some of the Bi$_2$Se$_3$ nanoribbon Josephson junctions. Inset shows enlarged conductance curve (dI/dV) for device B45-C1, (Table~S1, SI).} 
    \label{dI/dV}
\end{figure}
\indent The as grown nanoribbons \cite{Andzane2015a,Kunakova2018} were transferred to SiO$_2$ (300nm)/Si substrates and 
electron beam lithography processing was used to define the electrodes separated by a distance ranging from 70 nm to 150 nm. Prior to the deposition of the electrodes the Bi$_2$Se$_3$ nanoribbon is etched by Ar$^+$ ion milling to remove the layer of native oxide. This step is followed by the evaporation of the bilayer Pt and Al (3nm/80nm). In our earlier works \cite{Galletti2014, Galletti2014a, Galletti2017, Charpentier2017}, by using TI flakes, we have demonstrated the crucial role of Pt to create a transparent interface when Al is employed as electrode. This has been also confirmed by other authors \cite{Ghatak2018a}. All the measurements  presented  here were conducted in an $\it{rf}$-filtered dilution refrigerator with a base temperature of 19 mK.\\
\indent Fig.~\ref{DeviceSchem}a shows a Scanning Electron Microscope (SEM) image of one of the fabricated planar Josephson junctions of Al/Bi$_2$Se$_3$/Al. 
Ideally, a planar Josephson junction can be described as SIS$^\prime$I$^\prime$-N-I$^\prime$S$^\prime$IS (see schematics in Fig.~\ref{DeviceSchem}b). Here S$^\prime$ is the proximized TI (under the superconducting electrode S), characterized by an induced gap $\Delta^\prime$ while N is the part of the TI in the nanogap not covered by the superconductor (Fig.~\ref{DeviceSchem}b). I indicates the barrier between the Al and the Bi$_2$Se$_3$ nanoribbon while I$^\prime$ is the barrier between the Bi$_2$Se$_3$ under the Al and that Bi$_2$Se$_3$ in the nanogap.\\
\indent The current-voltage characteristic (IVC) of one of the nanoribbon Josephson junctions is shown in Fig.~\ref{DeviceSchem}c. The IVCs of our junctions typically display a hysteretic behaviour which is an indication of an increased electron temperature once the junction switches to the resistive state \cite{Courtois2008}. The switching current I$_c$ is determined from the forward sweep, and for the junction B45-C1 I$_c$ is 0.20 $\mu$A  (Table~S1, in Supplementary Information (SI)). This corresponds to a current density J$_c$ of 2.9$\times$10$^3$~A/cm$^2$. The value of the current density is comparable to other previously reported 3D TI Josephson junctions with much larger cross sections\cite{Galletti2014a, Stehno2016}.\\
\indent An important figure of merit of a Josephson junction is the characteristic voltage or I$_c$R$_N$ product. The normal state resistance R$_N$ can be extracted from the inverse of the slope of the IVC at voltages above 2$\Delta_{Al}$ ($\sim$320~$\mu$eV). For the junction B45-C1, R$_N$ of 330 $\Omega$ yields I$_c$R$_N$ = 66~$\mu$eV (Table~S1, SI). The I$_c$R$_N$ increases to about 170 $\mu$eV as increasing the width of the nanoribbon by a factor of two (see in Table~S1, SI).\\
\indent A common feature for all the investigated devices is the presence of an excess current (I$_{exc}$) in the IVC, which occurs due to Andreev reflection. I$_{exc}$ is obtained by extrapolating the linear part of the IVC, taken at voltages much higher than the gap of Al (see Fig.~\ref{DeviceSchem}c), to V~=~0. 
In the theoretical model first developed by Blonder, Tinkham and Klapwijk and later on modified by Flensberg,\cite{Flensberg1988} the transparency of $\tau$ the barrier I$^\prime$ in a S$^\prime$I$^\prime$NI$^\prime$S$^\prime$ junction can be related to dimensionless parameter Z which give the strength of barrier: one gets that $\tau$~=~1/(1~+~Z$^2$). The transparency $\tau$ is then connected to the value of the excess current through the quantity eI$_{exc}$R$_N$/$\Delta^{\prime}$, where R$_N$ is the resistance of the normal state.\\ 
 \begin{figure*}
    \centering
    \includegraphics[width=1.0\linewidth]{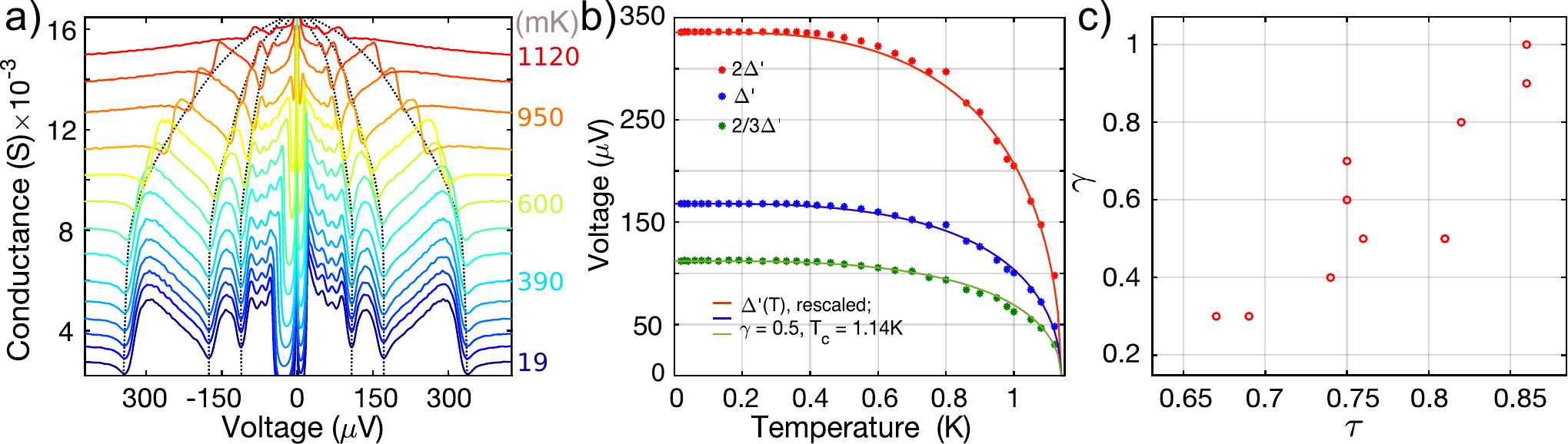}
    \caption{(a) (dI/dV) as a function of bias voltage for junction B45-C1 measured at various temperatures. 
(b) Temperature dependence of the first three dips in the conductance spectra n~=~1;~2;~3 (corresponding to 2$\Delta^\prime$; $\Delta^\prime$; 2/3$\Delta^\prime$). The solid red, blue and green curves are the calculated temperature dependences of the induced gap $\Delta^\prime$(T) in agreement with eq.~\ref{DeltaT}, considering the parameter $\gamma$~=~0.5 and that the T$_c$ of the junction is 1.14~K (Table~S1, SI). The calculated curves are scaled to the experimentally determined  $\Delta^\prime$ for n~=~2. (c) The parameter $\gamma$ extracted from the $\Delta^\prime$(T) curves 
for the various junctions considering eq.~\ref{DeltaT}, is plotted as a function of the transparency $\tau$.}    
    \label{TempD}
\end{figure*} 
\indent In Fig.~\ref{dI/dV}a we show the conductance spectra of some of the measured devices with slightly different geometrical parameters (L) $\times$ (w) $\times$ (t), (Table~S1, SI). The conductance spectra for all the junctions exhibit clear features at voltages below the gap of Al. 
In case of a very high transparency of the Josephson junction, the dips in the conductance spectra are associated to Multiple Andreev Reflections, occurring at voltages V=2$\Delta^{\prime}$/en, where n is an integer \cite{Kleinsasser1994,Averin1995}. Fig.~\ref{dI/dV}b depicts the enlarged conductance spectra for device B45-C1. The red arrows indicate the positions of the dips, that 
follow the relation 2$\Delta^{\prime}$/en. The induced gap $\Delta^{\prime}$ for this junction is found to be 168~$\mu$eV (Table~S1, SI). This value of the induced gap can 
be used to evaluate the transparency $\tau$, which for all junctions is found in the range 0.65~-~0.85, indicating a good transmission through the barrier I$^{\prime}$ (see Fig.~\ref{DeviceSchem}b). It is important to note, that such high transparency of the Josephson barrier interface is reproducible from nanoribbon to nanoribbon (Table~S1, SI).\\
\indent Fig.~\ref{TempD}a illustrates the conductance spectra for device B45-C1 measured at different temperatures. The temperature dependence of the induced gap $\Delta^{\prime}$ can be described as \cite{Aminov1996,Chrestin1997}:
 \begin{equation}
\Delta^\prime(T) = \dfrac{\Delta_{Al}}{1+\gamma\sqrt{\Delta_{Al}^2(T)-\Delta^{\prime 2}(T)}/k_B T_c}.
\label{DeltaT}
 \end{equation}
Here the parameter $\gamma$ is proportional to the ratio $R_B/\rho_N\xi_N$ where $\rho_N$ and $\xi_N$ are respectively the resistivity and the coherence length of the Bi$_2$Se$_3$ and $R_B$ the interface resistivity between Al and Bi$_2$Se$_3$.
The three main conductance dips that we observe (n = 1, 2, 3) are identified as 2$\Delta^\prime$, $\Delta^\prime$ and 2/3$\Delta^\prime$. They are plotted as a function of the temperature (Fig.~\ref{TempD}b) and can be well approximated using equation~\ref{DeltaT}. We find the best fit of the data $\Delta^\prime$(T) to eq.~\ref{DeltaT} by using $\gamma$~=~0.5 and the critical temperature of the junction T$_c$~=~1.14~K. The other two curves of the fit for the temperature dependence of the conductance dips 2$\Delta^\prime$(T) and 2/3$\Delta^\prime$(T) (red and green solid lines in Fig.~\ref{TempD}b) are obtained by rescaling to the $\Delta^\prime$(T) values (multiplying by 2 and 2/3 respectively) and keeping the same parameters of T$_c$ and $\gamma$. The good fitting of the 2$\Delta^\prime$/n curves to eq.~\ref{DeltaT} supports the assumption that the dips in the conductance spectra are due to the proximity induced superconducting gap in the Bi$_2$Se$_3$.\\
\indent Fig.~\ref{TempD}c shows the dependence of the extracted $\gamma$ values, for the junctions reported in Table~S1, as a function of the transparency of the barrier I$^\prime$ (extracted  from I$_{exc}$ \cite{Flensberg1988}). 
Quite surprisingly the lowest values of $\tau$ correspond to the lowest value of $\gamma$ meaning the transparency of the barrier I$^\prime$ is reduced by reducing the ratio $R_B/\rho_N\xi_N$, which corresponds to a larger value of the induces gap $\Delta^\prime$. This behaviour can be understood considering that the Al electrode material is also doping the Bi$_2$Se$_3$ underneath. A change in the transparency of the Al/Bi$_2$Se$_3$ interface (variation of $R_B$)  produces a change in the Bi$_2$Se$_3$ material, through doping, which modifies the product $\rho_N\xi_N$ as well. Our measurements show that the reduction in $R_B$ does not reflect  in an equal reduction of the $\rho_N\xi_N$ product, otherwise we would have not detected an increase in $\Delta^\prime$. The ``response'' of the product $\rho_N\xi_N$ to variation of $R_B$ will depend on the initial doping of the material and it can eventually be tuned by gating. The improved coupling between the Al and the Bi$_2$Se$_3$  leads to a more doped TI underneath the electrodes, which  naturally causes a difference between the chemical potentials of the Bi$_2$Se$_3$  doped by the Al electrodes compared to   the Bi$_2$Se$_3$ in the nanogap. The resulting Fermi momentum mismatch leads to an effective barrier I$^\prime$, which will be more or less transparent depending on the extrinsic doping level difference (coming from the Al electrodes) in the two Bi$_2$Se$_3$ regions.
While high transparent Bi$_2$Se$_3$ / Al interfaces are instrumental to get high values of the induced gap important for high working temperature of the device, this happens at the expenses of the transparency of the interface I$^\prime$ which affects the visibility of the (Majorana) Andreev bound states leading to a 4$\pi$ periodic CPR. Indeed the number of modes leading to a 4$\pi$ periodic CPR increases for higher values of the transparency of the barrier I$^\prime$ \cite{Snelder2013,Li2018a}. A precise control of the interface between the Al and the Bi$_2$Se$_3$ is therefore important to reach values of $\tau$ higher than 0.85 achieved in this work.\\ 
\indent To further corroborate the quality of our 3D TI junctions we have performed low frequency resistance noise measurements at voltages above 2$\Delta^\prime$/e. The resistance fluctuations $\delta$R are measured by applying a constant bias current and recording the voltage fluctuations as a function of time $\delta$V(t)~=~$\delta$R(t)~I$_b$. The time dependent voltage signal is then amplified by a room temperature low noise amplifier and the output sent to an FFT spectrum analyser. The amplifier noise contribution was further reduced by using two amplifiers in  parallel and performing a cross-correlation spectrum in the FFT analyser\cite{Arpaia2016}.\\
 \begin{figure}
    \centering
    \includegraphics[width=0.5\linewidth]{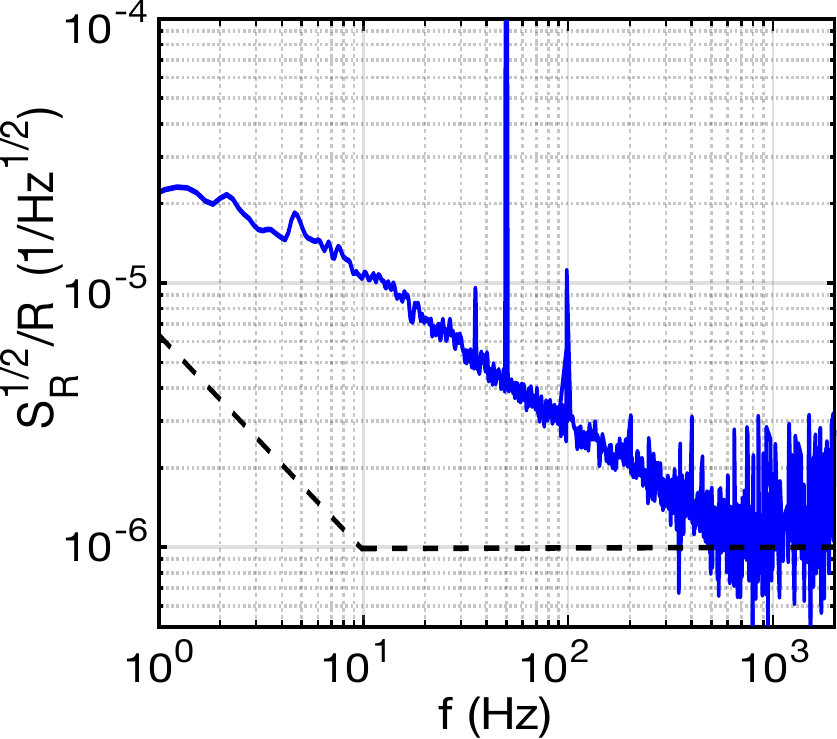}
    \caption{Normalized resistance noise power spectrum as a function of frequency measured at T~=~500~mK and bias voltage V~=~1.96~mV (solid blue line). The dashed line indicates the measurement setup background noise level.}    
    \label{Rnoise}
\end{figure} 
\indent In Fig.~\ref{Rnoise} we show the normalized resistance noise power spectral density, S$_R^{1/2}$/R~=~$\sqrt{\langle \delta R^2 \rangle}$/R, of junction B45-32 measured at T~=~500~mK. At frequencies $f$ below 1~kHz we clearly observe the typical 1/$f$ dependence of resistance noise common to solid state electronic systems \cite{Kogan1996}. Resistance fluctuations can be attributed to fluctuations of the transmission coefficients of the transport modes or the number of transport modes on the TI surface. An empirical equation for 1/$f$ resistance noise is given by Hooge's expression for low frequency voltage noise S$_V$/V$^2$~=~S$_R$/R$^2$~=~$\alpha_H$/($f n A$), where $f$ is the frequency, $n$ is the carrier density, $A$ is the surface area of the TI between the superconducting electrodes, and $\alpha_H$ is the Hooge's parameter, which is a measure for the "noisiness" of the system. Using a typical total carrier concentration of 1$\times$10$^{13}$~cm$^{-2}$ for our nanoribbons \cite{Kunakova2018} we obtain a Hooge's parameter $\alpha_H$~=~7$\times$10$^{-7}$. We have measured low frequency noise on 3 more TI junctions, which resulted in a Hooge's parameter in the range 5~$-$~9$\times$10$^{-7}$. These values are 3 to 4 orders of magnitude lower than those reported in literature on Bi based TI junctions measured at T~=~4~$-$~7~K \cite{Zhang2016, Bhattacharyya2015}. The low noise values can be attributed to the quasi ballistic limit of our 3D TI junctions and the very clean interfaces in our devices resulting in 
low fluctuations of the transport mode transmission coefficients.\\
\indent To identify the microscopic origin of the resistance noise observed in our 3D TI junctions a detailed study of noise as a function of temperature and external gate voltage will be required \cite{Zhang2016, Bhattacharyya2015}. However, this is beyond the scope of this paper. 
Resistance noise in a Josephson junction gives a lower bound for critical current noise, which is a source of dephasing in Josephson junction based superconducting qubits \cite{VanHarlingen2004}. For experiments aiming at detecting Majorana zero energy modes in TI Josephson junctions embedded in circuit quantum electrodynamics architectures \cite{Ginossar2014} qubit we compare the resistance noise of our TI junctions to Al tunnel junctions employed in state of the art Josephson junction qubit architectures. Typical reported values of the relative resistance noise at 1~Hz multiplied by the junction area A$_J$ (perpendicular to the transport current), i.e. A$_J$S$_R$/R$^2$, are in the range 2$\times$10$^{-14}$~$-$~2$\times$~10$^{-13}$~$\mu$m$^2$/Hz \cite{Eroms2006a, Nugroho2013}. For our junctions we obtain values in the range 3$\times$10$^{-13}$~$-$~1$\times$~10$^{-12}$~$\mu$m$^2$/Hz, which are close to the best reported values for Al tunnel junctions.\\
\indent In conclusion we have realized low noise and high transparency 3D TI nanoribbon Josephson junctions 
instrumental to study topological superconductivity in few modes devices.
\section*{Acknowledgements}
This work has been supported by the European Union's Horizon 2020 research and innovation programme (grant agreement No. 766714/HiTIMe). G.K. acknowledges European Regional Development Fund project No 1.1.1.2/VIAA/1/16/198.
\section*{Supplementary Information}
See Supplementary Information (SI) for physical parameters of the studied junctions, extraction of the induced gap  $\Delta^\prime$ from conductance vs voltage measurements, and resistance vs. temperature measurements of a junction.


\begin{thebibliography}{widest entry}
\bibitem{Fu2008}
L. Fu and C. L. Kane, ``Superconducting proximity effect and majorana fermions at the surface of a topological insulator", Physical Review Letters 100, 096407 (2008).

\bibitem{Wiedenmann2016}
J. Wiedenmann, E. Bocquillon, R. S. Deacon, S. Hartinger, O. Herrmann, T. M. Klapwijk, L. Maier, C. Ames, C. Brune, C. Gould, A. Oiwa, K. Ishibashi, S. Tarucha, H. Buhmann, and L. W. Molenkamp, ``4$\pi$-periodic Josephson supercurrent in HgTe-based topological Josephson junctions", Nature Communications 7, 10303 (2016).

\bibitem{Snelder2013}
M. Snelder, M. Veldhorst, A. A. Golubov, and A. Brinkman, ``Andreev bound states and current-phase relations in three - dimensional topological insulators", Physical Review B - Condensed Matter and Materials Physics 87, 1–7 (2013).

\bibitem{Li2018a}
C. Li, J. C. de Boer, B. de Ronde, S. V. Ramankutty, E. van Heumen, Y. Huang, A. de Visser, A. A. Golubov, M. S. Golden, and A. Brinkman, ``4$\pi$-periodic Andreev bound states in a Dirac semimetal", Nature Materials 17, 875–880 (2018).

\bibitem{Manousakis2017}
J. Manousakis, A. Altland, D. Bagrets, R. Egger, and Y. Ando, ``Majorana qubits in a topological insulator nanoribbon architecture", Physical Review B - Condensed Matter and Materials Physics 95, 165424 (2017).

\bibitem{Chrestin1997}
A. Chrestin, T. Matsuyama, and U. Merkt, ``Evidence for a proximity-induced energy gap in Nb / InAs / Nb junctions", Physical Review B - Condensed Matter and Materials Physics 55, 8457–8465 (1997).

\bibitem{Kjaergaard2017}
M. Kjaergaard, H. J. Suominen, M. P. Nowak, A. R. Akhmerov, J. Shabani, C. J. Palmstr$\o$m, F. Nichele, and C. M. Marcus, ``Transparent Semiconductor-Superconductor Interface and Induced Gap in an Epitaxial Heterostructure Josephson Junction", Phys. Rev. Appl., 034029 (2017).

\bibitem{Chang2015}
W. Chang, S. M. Albrecht, T. S. Jespersen, F. Kuemmeth, P. Krogstrup, J. Nyg{\aa}rd, and C. M. Marcus, ``Hard gap in epitaxial semiconductor-superconductor nanowires", Nature Nanotechnology 10, 232–236 (2015).

\bibitem{Cho2013}
S. Cho, B. Dellabetta, A. Yang, J. Schneeloch, Z. Xu, T. Valla, G. Gu, M. J. Gilbert, and N. Mason, ``Symmetry protected Josephson supercurrents in three-dimensional topological insulators", Nature Communications 4, 1686–1689 (2013).

\bibitem{Galletti2014}
L. Galletti, S. Charpentier, M. Iavarone, P. Lucignano, D. Massarotti, R. Arpaia, Y. Suzuki, K. Kadowaki, T. Bauch, A. Tagliacozzo, F. Tafuri, and F. Lombardi, ``Influence of topological edge states on the properties of Al/Bi$_2$Se$_3$/Al hybrid Josephson devices", Physical Review B - Condensed Matter and Materials Physics 89, 134512–9 (2014).

\bibitem{Calvez2018}
K. L. Calvez, L. Veyrat, F. Gay, P. Plaindoux, C. Winkelmann, H. Courtois, and B. Sac{\`e}p{\`e}, ``Joule overheating poisons the fractional ac Josephson effect in topological Josephson junctions", Communications Physics , 1–9 (2018).

\bibitem{Kong2011a}
D. Kong, J. J. Cha, K. Lai, H. Peng, J. G. Analytis, S. Meister, Y. Chen, H. J. Zhang, I. R. Fisher, Z. X. Shen, and Y. Cui, ``Rapid surface oxidation as a source of surface degradation factor for Bi$_2$Se$_3$", ACS Nano 5, 4698–4703 (2011).

\bibitem{Andzane2015a}
J. Andzane, G. Kunakova, S. Charpentier, V. Hrkac, L. Kienle, M. Baitimirova, T. Bauch, F. Lombardi, and D. Erts, ``Catalyst-free vapour-solid technique for deposition of Bi$_2$Te$_3$ and Bi$_2$Se$_3$ nanowires/nanobelts with topological insulator properties", Nanoscale 7, 15935 (2015).

\bibitem{Kunakova2018}
G. Kunakova, L. Galletti, S. Charpentier, J. Andzane, D. Erts, F. L{\`e}onard, C. D. Spataru, T. Bauch, and F. Lombardi, ``Bulk- Free Topological Insulator Bi$_2$Se$_3$ nanoribbons with Magnetotransport Signatures of Dirac Surface States", Nanoscale 10, 19595–19602 (2018).

\bibitem{Galletti2014a}
L. Galletti, S. Charpentier, P. Lucignano, D. Massarotti, R. Arpaia, F. Tafuri, T. Bauch, Y. Suzuki, A. Tagliacozzo, K. Kadowaki, and F. Lombardi, ``Josephson effect in Al/Bi$_2$Se$_3$/Al coplanar hybrid devices", Physica C: Superconductivity 503, 162–165 (2014).

\bibitem{Galletti2017}
L. Galletti, S. Charpentier, Y. Song, D. Golubev, S. M. Wang, T. Bauch, and F. Lombardi, ``High-Transparency Al/Bi$_2$Te$_3$ Double-Barrier Heterostructures", IEEE Transactions on Applied Superconductivity 27, 1800404–4 (2017).

\bibitem{Charpentier2017}
S. Charpentier, L. Galletti, G. Kunakova, R. Arpaia, Y. Song, R. Baghdadi, S. M. Wang, A. Kalaboukhov, E. Olsson, F. Tafuri, D. Golubev, J. Linder, T. Bauch, and F. Lombardi, ``Induced unconventional superconductivity on the surface states of Bi$_2$Te$_3$ topological insulator", Nature Communications 8, 2019 (2017).

\bibitem{Ghatak2018a}
S. Ghatak, O. Breunig, F. Yang, Z. Wang, A. A. Taskin, and Y. Ando, ``Anomalous Fraunhofer Patterns in Gated Josephson Junctions Based on the Bulk-Insulating Topological Insulator BiSbTeSe$_2$", Nano Letters 18, 5124–5131 (2018).

\bibitem{Courtois2008}
H. Courtois, M. Meschke, J. T. Peltonen, and J. P. Pekola, ``Origin of hysteresis in a proximity Josephson junction", Physical Review Letters 101, 1–4 (2008).

\bibitem{Stehno2016}
M. P. Stehno, V. Orlyanchik, C. D. Nugroho, P. Ghaemi, M. Brahlek, N. Koirala, S. Oh, and D. J. Van Harlingen, ``Signature of a topological phase transition in the Josephson supercurrent through a topological insulator", Physical Review B - Condensed Matter and Materials Physics 93, 1–10 (2016).

\bibitem{Flensberg1988}
K. Flensberg, J. B. Hansen, and M. Octavio, ``Subharmonic energy gap structure in superconducting weak links", Physical Review B - Condensed Matter and Materials Physics 38, 8707– 8711 (1988).

\bibitem{Kleinsasser1994}
A. W. Kleinsasser, R. E. Miller, W. H. Mallison, and G. B. Arnold, ``Observation of Multiple Andreev Reflections in Superconducting Tunnel Junctions", Physical Review Letters 72, 1738–4 (1994).

\bibitem{Averin1995}
D. Averin and A. Bardas, ``ac Josephson effect in a single quantum channel", Physical Review Letters 75, 1831–1834 (1995).

\bibitem{Aminov1996}
B. A. Aminov, A. A. Golubov, and M. Y. Kupriyanov, ``Quasiparticle current in ballistic constrictions with finite transparencies of interfaces", Phys. Rev. B 53, 365–373 (1996).

\bibitem{Arpaia2016}
R. Arpaia, M. Arzeo, S. Nawaz, S. Charpentier, F. Lombardi, and T. Bauch, ``Ultra low noise YBa$_2$Cu$_3$O$_{7-\delta}$ nano superconducting quantum interference devices implementing nanowires", Applied Physics Letters 252601, 072603–4 (2016).

\bibitem{Kogan1996}
S. Kogan, Electronic Noise and Fluctuations in Solids, Cambridge University Press, Cambridge, 1996.

\bibitem{Zhang2016}
H. Zhang, Z.-J. Song, J.-Y. Feng, Z.-Q. Ji, and L. Lu, ``Low Frequency Noise in Gate Tunable Topological Insulator Nanowire Field Emission Transistor near the Dirac Point", Chinese Physics Letters 33, 087302 (2016).

\bibitem{Bhattacharyya2015}
S. Bhattacharyya, M. Banerjee, H. Nhalil, S. Islam, C. Dasgupta, S. Elizabeth, and A. Ghosh, ``Bulk-Induced 1/f Noise at the Surface of Three-Dimensional Topological Insulators", ACS Nano 9, 12529–12536 (2015).

\bibitem{VanHarlingen2004}
D. J. Van Harlingen, T. L. Robertson, B. L. T. Plourde, P. A. Reichardt, T. A. Crane, and J. Clarke, ``Decoherence in josephson junction qubits due to critical current fluctuations", Phys. Rev. B 70, 064517 (2004).

\bibitem{Ginossar2014}
E. Ginossar and E. Grosfeld, ``Microwave transitions as a signature of coherent parity mixing effects in the majorana transmon qubit", Nature Communications 5, 4772 (2014).

\bibitem{Eroms2006a}
J. Eroms, L. C. van Schaarenburg, E. F. C. Driessen, J. H. Plantenberg, C. M. Huizinga, R. N. Schouten, A. H. Verbruggen, C. J. P. M. Harmans, and J. E. Mooij, ``Low-frequency noise in Josephson junctions for superconducting qubits", Applied Physics Letters 89, 122516 (2006).

\bibitem{Nugroho2013}
C. D. Nugroho, V. Orlyanchik, and D. J. Van Harlingen, ``Low frequency resistance and critical current fluctuations in Al-based Josephson junctions", Applied Physics Letters 102, 142602 (2013).

 \end{thebibliography}

\end{document}


\title {High transparency Bi$_2$Se$_3$ topological insulator nanoribbon Josephson junctions with low resistive noise properties: Supplementary Information}

\author{Gunta Kunakova}
\affiliation{Quantum Device Physics Laboratory, Department of Microtechnology and Nanoscience, Chalmers University of Technology, SE-41296 G\"oteborg, Sweden}
\affiliation{Institute of Chemical Physics, University of Latvia, Raina Blvd. 19, LV-1586, Riga, Latvia}

\author{Thilo Bauch}
\affiliation{Quantum Device Physics Laboratory, Department of Microtechnology and Nanoscience, Chalmers University of Technology, SE-41296 G\"oteborg, Sweden}

\author{Edoardo Trabaldo}
\affiliation{Quantum Device Physics Laboratory, Department of Microtechnology and Nanoscience, Chalmers University of Technology, SE-41296 G\"oteborg, Sweden}

\author{Jana Andzane}
\affiliation{Institute of Chemical Physics, University of Latvia, Raina Blvd. 19, LV-1586, Riga, Latvia}

\author{Donats Erts}
\affiliation{Institute of Chemical Physics, University of Latvia, Raina Blvd. 19, LV-1586, Riga, Latvia}

\author{Floriana Lombardi}
\email{floriana.lombardi@chalmers.se}
\affiliation{Quantum Device Physics Laboratory, Department of Microtechnology and Nanoscience, Chalmers University of Technology, SE-41296 G\"oteborg, Sweden}

\date{\today}

\maketitle
\section*{}
\section{Geometry and transport parameters of Al/Bi$_2$Se$_3$ nanoribbon /Al Josephson junctions}

\begin{table}[h!] 
\renewcommand{\thetable}{S\arabic{table}} 
\caption{Geometric and transport parameters of the Al/Bi$_2$Se$_3$ nanoribbon /Al Josephson junctions.}
  \label{tbl:Summary}
\begin{tabular}{lcccccccc}
\noalign{\smallskip}\hline\noalign{\smallskip}
Junct. & 
\begin{tabular}{c}L$\times$w$\times$t\emph{*}, (nm)\end{tabular}&
\begin{tabular}{c}I$_c$, ($\mu$A)\end{tabular}&
\begin{tabular}{c}I$_c$R$_N$, ($\mu$eV)\end{tabular}&
\begin{tabular}{c}I$_{exc}$, ($\mu$A) \end{tabular}&
$\tau$ & 
\begin{tabular}{c}$\Delta^\prime$, ($\mu$eV) \end{tabular}&
$\gamma$\\
\hline\noalign{\smallskip}
B45-C1 &100$\times$430$\times$16 &0.20 &66 &0.63 &0.81 &168 & 0.5\textsuperscript{\emph{a}}\\
B53-112 &100$\times$150$\times$14 &0.12 &58 &0.34 &0.75 &167 & 0.6\textsuperscript{\emph{a}}\\
B53-123 &80$\times$270$\times$34 &1.04 &144 &1.50 &0.82 &160 & 0.8\textsuperscript{\emph{c}}\\
B53-91 &80$\times$870$\times$14 &2.20 &152 &3.50 &0.86 &160 & 0.9\textsuperscript{\emph{d}}\\
B45-C2 &70$\times$430$\times$16 &0.36 &82 &0.92 &0.81 &170 & 0.5\textsuperscript{\emph{a}}\\
B45-D4 &70$\times$60$\times$20 &0.09 &59 &0.32 &0.81 &163 & -\\
B53-5 &135$\times$215$\times$27 &0.30 &82 &0.61 &0.75 &166 & 0.7\textsuperscript{\emph{a}}\\
B53-92 &50$\times$870$\times$14 &3.40 &180 &4.50 &0.86 &157 & 1.0\textsuperscript{\emph{d}}\\
B45-32 &100$\times$70$\times$20 &0.025 &34 &0.12 &0.74 &168 & 0.4\textsuperscript{\emph{a}}\\
B24-9R1 &100$\times$400$\times$74 &1.53 &101 &2.80 &0.79 &160 & -\\
B45-A3 &100$\times$160$\times$38 &0.092 &52 &0.31 &0.76 &167 & 0.5\textsuperscript{\emph{b}}\\
B45-A2/1 &180$\times$70$\times$15 &0.011 &19 &0.07 &0.67 &162 & 0.3\textsuperscript{\emph{a}}\\
B45-B4 &70$\times$60$\times$13 &0.010 &23 &0.06 &0.69 &176 & 0.3\textsuperscript{\emph{a}}\\
\hline
\end{tabular}

\textsuperscript{\emph{*} L - length, w - width, t - thickness.}
\textsuperscript{\emph{a; b; c; d}: T$_c$= a) 1.14K; b) 1.13K; c) 1.11K; d) 1.08K}\\

\label{Tab}
\end{table}

\newpage

\section{Extraction of induced gap $\Delta^\prime$}
\begin{figure*}[h!]
\renewcommand{\thefigure}{S\arabic{figure}} 
    \centering
    \includegraphics[width=0.95\linewidth]{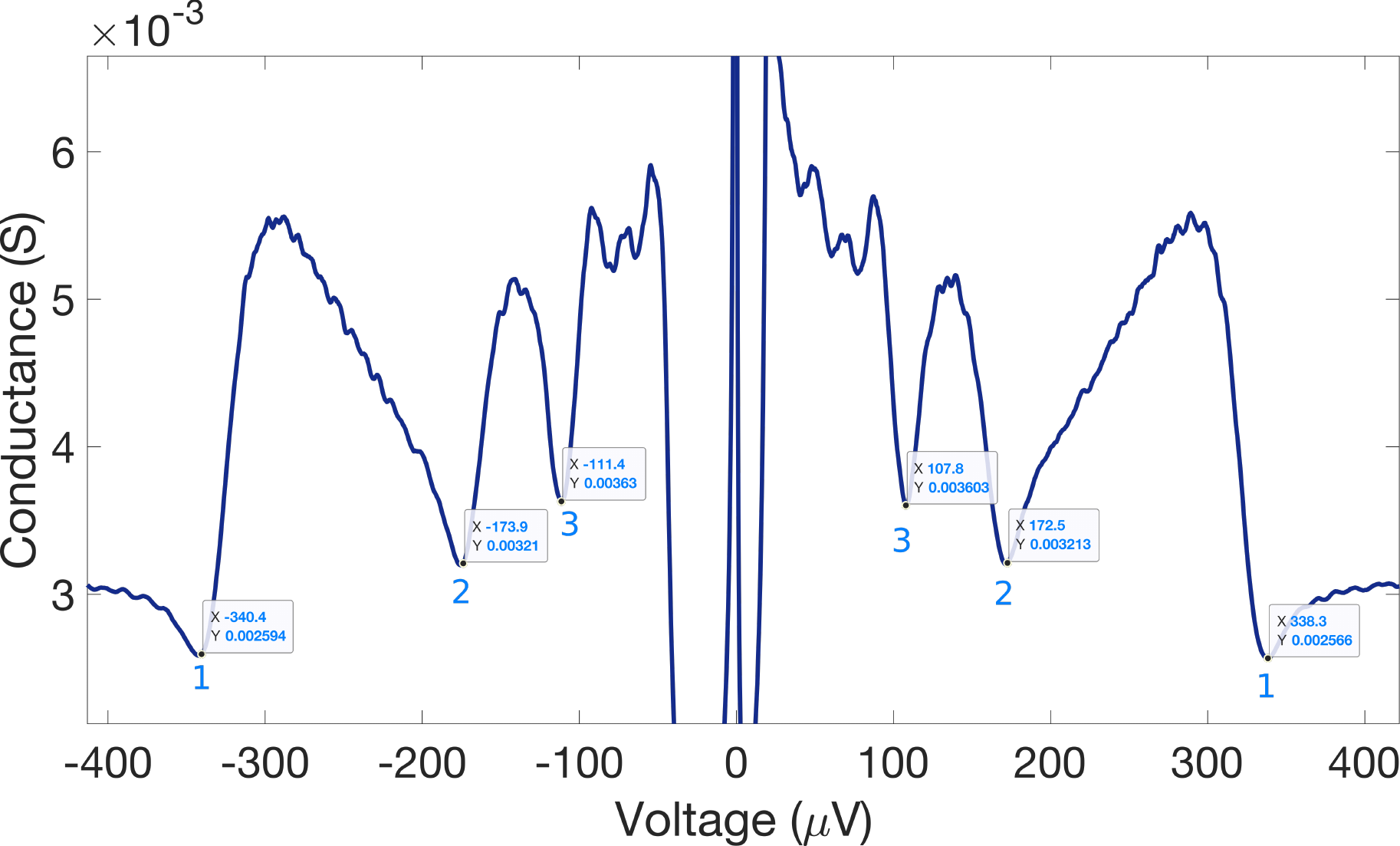}
    \caption{Conductance spectrum of Bi$_2$Se$_3$ nanoribbon Josephson junction B45-C1. For a highly transparent interfaces, the dips in the conductance vs. voltage curve, are due to Multiple Andreev reflection; by indexing them with integer numbers one can retrieve the value of the induced gap $\Delta^\prime$. In our data, the voltages corresponding to three most pronounced dips in the conductance curve were extracted for positive and negative voltages (see markers).  The extracted values for both positive and negative voltages were averaged and $\Delta^\prime$ was calculated for each value of  $n$ as $2\Delta^\prime /n$. As a result, three values of $\Delta^\prime$ were obtained (for each dip) which were then averaged yielding the final value of $\Delta^\prime$ (168~$\mu$V for junction B45-C1, listed in Supplementary Table S1). This procedure to extract the $\Delta^\prime$ was repeated also for other junction listed in Table S1.}
    \label{Fig:S1}
\end{figure*}

\newpage

\section{Resistance vs. temperature}

\begin{figure*}[h!]
\renewcommand{\thefigure}{S\arabic{figure}} 
    \centering
    \includegraphics[width=0.95\linewidth]{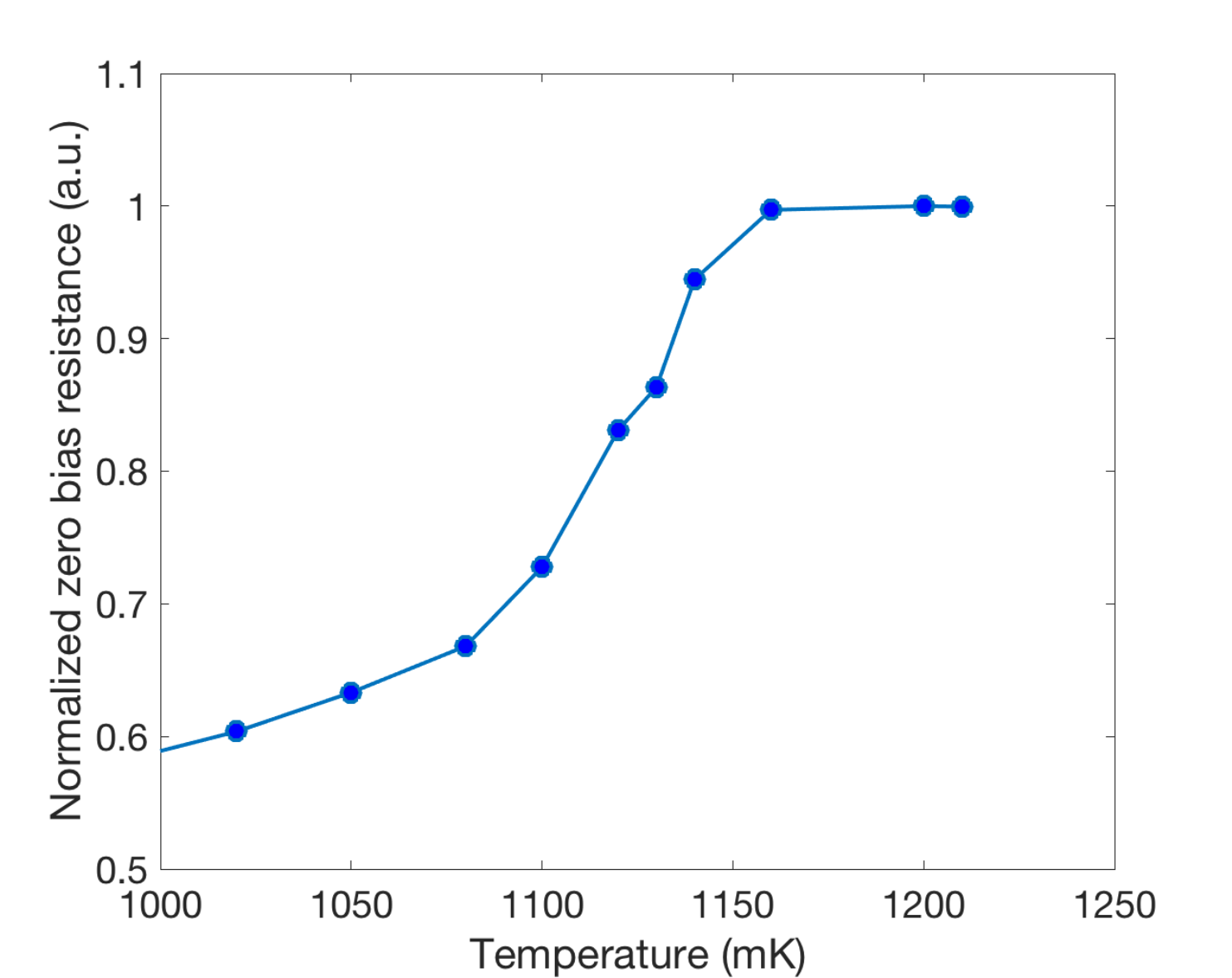}
    \caption{Zero bias resistance vs temperature of junction B45-B4. At temperatures above 1150 mK the resistance stays constant corresponding to the normal state of the Al electrodes. For temperatures $\leq 1140$~mK the resistance drops monotonically for decreasing temperatures. We define the onset of the resistive transition in temperature $T=1140$~mK as the critical temperature of the Al electrodes.}
    \label{Fig:S2}
\end{figure*}